\documentstyle[aps,twocolumn]{revtex}

\begin{document}
\draft

\title{The Schwarz-Hora effect: present-day situation}

\author{Yu. N. Morokov}
\address{
Institute of Computational Technologies, Siberian Branch of \\
the Russian Academy of Sciences, Novosibirsk 630090, Russia}

\maketitle

\begin{abstract}
The Schwarz-Hora effect, observed under attempts to modulate an 
electron beam by laser light, was discussed extensively in the 
literature in the early 1970s. The analysis of the literature 
shows there are the unresolved up to now contradictions between 
the theory and the Schwarz experiments. The new model for the 
interpretation of the Schwarz-Hora effect is proposed. The 
problem of the description of the long-wavelength spatial beating 
in the Schwarz-Hora radiation is essential for the model and is 
considered in detail. \\

\end{abstract}

\narrowtext

\section{INTRODUCTION} 

\label{sect:intro} 

\hspace {4.5mm}
Thirty years ago, H. Schwarz \cite{1} attempted to modulate an 
electron beam with optical frequency. In his experiments 
\cite{1,2,3,4,5,6}, a new effect has been discovered. When a 
50-keV electron beam inside an ultra-high-vacuum system crossed a 
thin crystalline dielectric film illuminated with laser light, 
electrons produced the electron-diffraction pattern not only at a 
fluorescent target but also at a nonfluorescent target (the 
Schwarz-Hora effect). In the latter case the pattern was roughly 
of the same color as the laser light. The effect was absent if 
the electrical vector of the polarized laser light was parallel 
to the film surface. When changing the distance between the thin 
crystalline film and the target, a periodic change in the light 
intensity was observed with spatial period of the order of 
centimeters. Since 1972 no reports on the results of further 
attempts to repeat those experiments in other groups have 
appeared, while the failures of the initial such attempts have 
been explained by Schwarz in Ref.\cite{5}. The latest review 
can be found in Ref.\cite{7}. 

The reported quantitative results \cite{2,3,4,5,6,7} were 
obtained for the films of about 1000 $\AA $ thickness. The main 
material used in the experiments was SiO$_2$ \cite{7}. The films 
were illuminated by a 10$^7$-W/cm$^2$ argon ion laser irradiation 
($\lambda _p$= 4880 $\AA $) perpendicular to the electron beam of 
about 0.4 $\mu $A current. These values will be used below for 
estimates.

There are several essential contradictions between the theory and 
the Schwarz experiments: 

1. {\it The radiation intensity.} The relatively high intensity 
of the Schwarz-Hora radiation (at least of the order of 
10$^{-10}$ W) was observed in the Schwarz experiments. The 
calculated power of the coherent emission of light at the laser 
frequency turns out to be at least 10$^3$ times smaller 
\cite{7,8,9,10,11,12,13,14}. 

2. {\it The dependence on the electron current.} The presented in 
Refs.\cite{2,3} experimental photographs of the diffraction 
pattern allow to affirm that the intensity of the Schwarz-Hora 
radiation is linear on an electron beam current \cite{15}. The 
quantum-mechanical treatment shows that a sharp peak in the 
intensity at the laser frequency can be accounted for only the 
collective processes of light emission 
\cite{8,9,10,11,12,13,16,17,18}. It leads to the quadratic 
dependence of the radiation on the electron current. 

3. {\it The polarization angle dependence.} The observed 
intensity of the Schwarz-Hora radiation falls off exponentially 
with the angle between the electric vector of the laser light and 
the electron beam direction \cite{3,7,10,19}. The exponential 
slope appears to be linear in the distance $z$ between the film 
surface and the target. A theoretical explanation of such 
dependence is absent. 

4. {\it An initial phase of the spatial beating.} The Schwarz 
experiments indicate \cite{3,19} that there must be the maximum 
of the Schwarz-Hora radiation intensity at the dielectric film 
surface. The quantum-mechanical models predict the minimum 
\cite{8,9,11,14,15,17,20,21}. 

5. {\it The spatial beating period.} There is the large 
discrepancy of more than 10$\%$ between the experimental and 
theoretical results for the period of the spatial beating of the 
Schwarz-Hora radiation \cite{21}. 

These contradictions allow to conclude that we have not the 
adequate theory for the interpretation of the Schwarz-Hora 
effect. It seems natural in this situation to try to connect the 
known facts in some phenomenological scheme. Below we consider 
such phenomenological model. The quantitative aspects of the 
model are essentially connected with the interpretation of the 
long-wavelength spatial modulation of the Schwarz-Hora radiation 
\cite{3,6,7,8,10,11,15,17,20,21,22,23}. The quantum 
interpretation of such modulation was discussed in 
Ref.\cite{21}.

\section{QUANTUM INTERPRETATION OF THE LONG-WAVELENGTH SPATIAL 
         BEATING}

\label{sect:beating1} 

\hspace {4mm}
Let the $z$ axis be directed along the incident electron beam. 
The laser beam is along the $x$ axis. The electrical vector of 
the laser light is in the $z$ direction. Electrons pass through 
the dielectric slab restricted by the planes $z = -d$ and 
$z = 0$. We consider without loss of generalization only the 
central outgoing electron beam (zeroth-order diffraction). 

Usually the following assumptions are used: An electron interacts 
with the light wave only within the slab; it interacts within the 
slab only with the light wave; the spin effects can be neglected. 
In the simplest case the light field within the slab and incident 
electrons are represented by plane waves. 

Using these assumptions, consider the origin of the 
long-wavelength spatial modulation in the one-electron quantum 
theory. The solution of the Klein-Gordon equation to first order 
in the light field (see, for example, Refs.\cite{8,20}) gives 
the following expression for the electron probability density for 
$z > 0$: 
\begin{eqnarray}
\label{1}
\rho (x,z,t) = \rho _0 \biggl\{ 1-\beta \sin \left[ 
\frac{z}{2\hbar } (2p_0-p_{1z}-p_{-1z})\right] \nonumber \\ 
\times \sin \left( \frac{\pi d}{2d_0}\right) 
\cos \left[ kx - \omega t+\frac{z}{2\hbar } 
(p_{1z}-p_{-1z})\right] \biggl\} .
\end{eqnarray}
Here $\rho _0$ is the probability density for the initial 
incident electron beam and $\omega $ and $k$ denote the circular 
frequency and the wave number of the light wave inside the slab. 
The parameter $\beta $ is proportional to the amplitude of the 
laser field and $d_0$ is the smallest optimum value of the slab 
thickness. For the conditions of the Schwarz experiments, these 
parameters are $\beta $= 0.35 (for $\alpha $-quartz) and $d_0$ = 
1007 $\AA $. This implies that the probability for absorption (or 
stimulated emission) of a photon by an electron inside the slab 
is $(\beta /4)^2$ = 0.008 for $d$ = $d_0$. The value $n_o = 
1.550$ was used here as the $\alpha $-quartz refractive index. 
The $z$ components of the momentum $p_{nz}$ are determined for 
free electrons of energy $E_n$ and momentum ${\rm\bf p}_n$ from 
the relativistic relationship 
\begin{eqnarray}
\label{2}
& & E_n^2 = m^2c^4 + {\rm\bf p}_n^2c^2, \\ 
& & E_n = E_0 + n\hbar \omega , \quad 
p_{nx} = n\hbar k, \quad 
n = 0,\pm 1. \nonumber 
\end{eqnarray}
Here $m$ is the electron mass. 

The probability that an electron absorbs or emits a photon inside 
the dielectric slab is a periodic function of the slab thickness. 
This is indicated by the second sine term in Eq.\ (\ref{1}). The 
experimental data on such dependence of the Schwarz-Hora 
radiation are absent in the literature. The cosine term 
represents the optical modulation of the electron beam. The 
first sine term in Eq.\ (\ref{1}) is a function of the distance 
$z$ between the slab and the target and represents the stationary 
modulation of the electron probability density. On equating the 
phase of this sine to 2$\pi z/\lambda _b$ (the same phase is 
obtained in the many-electron treatment \cite{10,11,17}) and 
taking into account the smallness of the ratio $\hbar \omega 
/E_0$, we obtain the expression for the spatial beating 
wavelength \cite{8}. 
\begin{eqnarray}
\label{3}
\lambda _b = \lambda _{b0}\frac{1}{1-(\frac {v_0}{c})^2(1-n^2)}, 
\end{eqnarray}
where $n = k c/\omega $ is the refractive index of the dielectric 
slab and 
\begin{eqnarray}
\label{4}
\lambda _{b0} = 2\lambda _p\left( \frac{E_0}{\hbar \omega }
\right) \left( \frac{v_0}{c}\right) ^3. 
\end{eqnarray}

The ratio of the initial electron velocity to the velocity of 
light in vacuum is $v_0/c$ = 0.4127 and $E_0/\hbar \omega = 
2.208\times 10^5$ for $E_0-mc^2$ = 50 keV. Then $\lambda _{b0} 
= 1.515$ cm. 

The following experimental values for $\lambda _b$ are reported: 
1.70 \cite{3}, 1.75 \cite{22}, and 1.73$\pm $0.01 \cite{23} cm. 
Equation (\ref{3}) gives $\lambda _b=1.22$ cm for $\alpha 
$-quartz. Thus the considered model does not give the agreement 
with experiment for $\lambda _b$. 

The situation, however, can be somewhat improved. As noted in 
Refs.\cite{15,24}, only one propagation mode of the light wave 
TM$_0$ can be excited within the slab under the experimental 
conditions considered. The corresponding wave field can be 
represented by a superposition of two traveling plane waves, 
propagating at angles $\pm \alpha $ to the $x$ axis. These waves 
turn one into another upon total internal reflection at the slab 
surfaces. The condition for the appearance of the next mode 
TM$_1$ can be written as $d > \lambda _p /2\sqrt{n^2-1}$. For 
$\alpha $-quartz it means $d > 2040\AA $. 

In case the light field is represented by one TM mode, the 
relativistic quantum-mechanical treatment can be carried out by 
analogy with the previous case. Such treatment leads to the same 
sine term for the stationary spatial modulation as that term in 
Eq.\ (\ref{1}). We obtain the following expression \cite{21}
\begin{eqnarray}
\label{5}
\lambda _b = \lambda _{b0}\frac{1}{1-(\frac {v_0}{c})^2
(1-n^2\cos ^2\alpha )}. 
\end{eqnarray}
This formula gives a better value for the spatial beating 
wavelength, $\lambda _b = 1.47$ cm, for $\alpha $-quartz. 
However, the condition for total internal reflection, $n \cos 
\alpha > 1$, limits the possibility to improve the agreement 
between the theory and experiment by using the formula (\ref{5}). 
This implies that $\lambda _b = \lambda _{b0}= 1.515$ cm is the 
upper limit, which cannot be exceeded by any formal optimization 
of the parameters $n$ and $d$. 

Thus the considered above quantum models cannot resolve the 
discrepancy of more than 10$\%$ between theory and experiment for 
the quantity $\lambda _b$. This statement remains valid even if 
we take into account some uncertainty of the published 
experimental data on the parameters $n$ and $d$. Below we 
consider the more general model situation, taking into account 
the possible electron beam divergence. 

\section{INFLUENCE OF THE ELECTRON BEAM DIVERGENCE ON THE 
         LONG-WAVELENGTH SPATIAL BEATING}

\label{sect:beating2} 

\hspace {4.5mm}
Let the incident electron beam is represented by a fragment of 
the spherical wave (Fig. 1), propagating from a focal point $F$. 
This point is placed at a distance $r$ from the slab surface 
$z=0$. We shall consider incident electrons moving at small 
angles $\beta $ to a beam optical axis FA. Below the label $0$ 
marks electrons passing through the film without energy change. 
The labels $1$ and $2$ mark electrons absorbing or emitting a 
photon inside the film. 

The film thickness $d$ gives the additional phase differences 
among the three outgoing electron beams. However, these phase 
differences are the same in the considered approximation for 
paraxial rays as for the plane incident electron wave and do not 
influence the spatial beating phase. We shall neglect below the 
film thickness and take $d=0$. 

Consider the interference among the electron waves at the point 
$A$ at the target (on the optical axis of the incident electron 
beam). The phases of the three electron waves at a moment $t$ can 
be written as 
\begin{eqnarray}
\label{6}
 & & \phi_0(A,t)=\frac{1}{\hbar}\left [p_0(z+r) -  E_0t\right ], 
\nonumber \\ 
 & & \phi_1(A,t)=\frac{1}{\hbar}\left [\bar{p}_{1z}r+p_{1z}z-E_0t-
\hbar\omega t\right ], \\
 & & \phi_2(A,t)=\frac{1}{\hbar}\left [\bar{p}_{2z}r+p_{2z}z-E_0t+
\hbar\omega t\right ], \nonumber
\end{eqnarray} 
where $\bar{p}_{1z}$ and $\bar{p}_{2z}$ are the $z$ components of 
the momenta of electrons for $z<0$; $p_{1z}$ and $p_{2z}$ are the 
same for $z>0$. 

The spatial beating phase is given by the next relationship 
\begin{eqnarray}
\label{7}
\chi = \frac{1}{2}(2\phi _0 -\phi_1 -\phi _2) = 
\frac{r}{2\hbar}(2p_0-\bar{p}_{1z}-\bar{p}_{2z}) \nonumber \\ 
+ \frac{z}{2\hbar}(2p_0-p_{1z}-p_{2z}). 
\end{eqnarray} 
Taking into account the smallness of $\hbar \omega /E_0$, we 
obtain 
\begin{eqnarray}
\label{8}
 & & \chi =  \frac{2\pi z}{\lambda_{b0}}
\left \{1 - \left (\frac{v_0}{c}\right )^2\left [1-(n \cos{\alpha })^2
\frac{r}{z+r}\right ]\right \}. 
\end{eqnarray} 

When $\chi (z)$ is a linear function of the distance $z$, the 
constant spatial beating wavelength $\lambda _b$ is determined by 
relationship $\chi = 2\pi z/\lambda _b$. In more general case, we 
define the spatial beating wavelength as $\lambda _b(z) = 
2\pi(d\chi/dz)^{-1}$. In the simplest case, $r(z)=const$, Eq.\
(\ref{8}) gives 
\begin{eqnarray}
\label{9}
& & \lambda _b(z) = \lambda _{b0}\frac{1}{1-\left (\frac 
{v_0}{c}\right )^2
\left [1-(n \cos{\alpha })^2\frac{r^2}{(z+r)^2}\right ]}, \\ 
& & r = const. \nonumber  
\end{eqnarray} 

One fragment of the experimental dependence of the radiation 
intensity on the distance $z$ has been depicted in 
Ref.\cite{3}. This fragment includes a maximum at $z=10.2$ 
cm. We used the formula (\ref{8}) to find the values of $r$ those 
give for $\alpha $-quartz the maxima of the functions $\sin 
^2\chi $ and $\cos ^2\chi $ at $z=10.2$ cm. The function $\sin 
^2\chi $ corresponds to the theoretical initial phase of the 
spatial beating, obtained in the quantum-mechanical models 
\cite{8,9,11,14,15,17,20}. The function $\cos ^2\chi $ 
corresponds to the initial phase considered in 
Refs.\cite{3,6,19}. 

The plots of the function $\lambda_b(z)$ are presented in Fig. 2 
for the three values of the parameter $m = \chi_0 /\pi$. Here 
$\chi_0$ is the value of $\chi(z)$ at $z = 10.2$ cm. The integer 
$m$ correspond to the maximum of $\cos ^2\chi $ at $z = 10.2$ cm 
and the half-integer $m$ correspond to the maximum of 
$\sin ^2\chi $. The three values: $r$ = 4.57, 10.08, and 22.13 cm 
have been obtained for $m = 12$, 12.5, and 13, correspondingly. 

Figure 2 shows the essential dependence of the spatial beating 
wavelength on the distance $z$ under fixed $r$. There are two 
limit values of $\lambda_b$, which do not depend on the 
parameters $n$ and $d$: $\lambda_b = 1.826$ cm and $\lambda_{b0} 
= 1.515$ cm. The first is an asymptotic value for $z \rightarrow 
\infty $ and the second is the upper limit for $z=0$.

The presented in Ref.\cite{3} experimental recording includes 
only three maxima and, therefore, does not allow to verify the 
dependence $\lambda_b $ on $z$. However, there are reported also 
the maxima at two other points: $z = 15.3$ cm and $z = 34.0$ cm. 
These data are in agreement with the constant value $\lambda_b = 
1.70 \AA$ and contradict the plots of $\lambda_b(z)$ presented in 
Fig. 2. This discrepancy can be explained only if the distances 
$r$ and $z$ are varied synchronously keeping fixed a factor 
$r/(z+r)$ in Eq.\ (\ref{8}). In this case we must assume that 
Schwarz varied also the focus position $r$ in adjusting the 
electron optics \cite{6} for each $z$ so that the ratio $r/z$ was 
kept up fixed. This assumption allows to explain the constancy of 
$\lambda_b = 1.70 \AA$ for $\alpha $-quartz taking $r= 4.55-4.57$ 
cm at $z = 10.2$ cm. However, in this case the problem of the 
initial phase of the spatial beating remains unresolved. The 
constant value of $r/z$ could be slightly different in different 
experiments. This could be a cause of some uncertainty of the 
experimental data on $\lambda _b$. 

\section{PHENOMENOLOGICAL MODEL FOR INTERPRETATION OF THE 
SCHWARZ-HORA EFFECT}

\label{sect:model} 

\hspace {4.5mm}
Thirty-year history of the Schwarz-Hora effect clearly shows that 
the modern formalism of the quantum electrodynamics cannot 
explain the reported by Schwarz experimental facts. To resolve 
finally this long-time-standing problem the new control 
experiments are necessary. In the absence of ones it seems 
reasonable to try to connect the known facts in some sufficiently 
simple phenomenological scheme. 

A starting idea for the proposed below phenomenological scheme 
may be considered as a literal reading of the expressions from 
the papers by Schwarz and Hora: "... electron beams are modulated 
at transmission of a laser-illuminated solid, transferring and 
generating photons at nonluminescent targets" \cite{22}; "This 
means that the electrons really "carried" to the screen photons 
"picked up" in the interaction region within the dielectric slab" 
\cite{3}. 
 
According to the existing quantum theory, an electron interacting 
with laser light in the presence of third body can absorb a 
photon. Such process we shall also call below "the total 
absorption of a photon". 

\noindent{\bf Supposition 1.} Suppose there can be another final 
result of capturing of a photon by an electron besides the total 
absorption. Suppose some electrons can capture a photon and form 
some metastable state in which the captured photon keeps its 
individuality. 

\noindent{\bf Supposition 2.} Suppose the total energy, momentum 
and mass of such electron metastable state are the same as for an 
electron which has absorbed a photon.  

We use here the term "metastability" to mark the instability of 
the electron state relative to the interaction between the 
metastable electron and a third body. Such interaction leads with 
a high probability to a release of the captured photon. At the 
same time, the conservation of energy-momentum forbids the 
emission of the captured photon in the absence of third bodies. 
The high instability of such states accounts for the difficulty 
of their experimental detection. From this point of view, the 
Schwarz experiments have given a unique possibility to observe a 
formation of the electron metastable states at the dielectric 
film surface and their decay at the target. The ultra-high vacuum 
in the Schwarz experiments allowed to save the metastable state 
on the way to the target. 

The simple estimate shows that the intensity of the Schwarz-Hora 
radiation of 10$^{-10}$ W can be reached if about 0.1\% of 
electrons in the beam take part in the transport of the captured 
photons to the target. So, the probability for the formation of 
the electron metastable state is comparable in magnitude with the 
probability 0.8\% considered in Sect.~\ref{sect:beating1} for the 
total absorption (or stimulated emission) of a photon by an 
electron inside the film. 

For further development of the model, we consider the spatial 
beating of the Schwarz-Hora radiation. 

In our scheme, the spatial oscillations of the radiation intensity 
is a consequence of the interference among light fields formed by 
the "released" photons. For that the captured photon must 
transfer to the target the information on a phase of the laser 
light field. 

The spatial oscillations cannot be obtained if we consider only 
the electron metastable states with energy $E_0+\hbar \omega$. To 
explain the experiment we must suppose an existing of the 
metastable electrons which have the different energy. Moreover, 
the amplitudes of those two electron waves must be comparable in 
magnitude to obtain the modulation depth ($\sim$ 85\%) observed 
in the experiment. 

\noindent{\bf Supposition 3.} Suppose the second beam of 
electrons in the metastable states consists of electrons with 
energy $E_0$.

Consider a process of stimulated emission of a photon by an 
electron in the laser field. We may consider this process as 
going through three steps: (i) a laser photon is captured by the 
electron; (ii) this photon stimulates a formation of a second 
(stimulated) photon on the electron; (iii) both the photons are 
emitted. To interpret the appearance of the second electron beam 
with captured photons we may suppose as a work hypothesis that 
these electrons appear as a result of the emission of only one 
photon at the third step (iii) of the stimulated emission 
process. If so, then the dielectric film surface gives a unique 
possibility for some electrons to "freeze" one of two stimulated 
photons.  

Consider the phases $\varphi_0$ and $\varphi_1$ of the light 
fields formed by the photons that were "thrown off" at the target 
by electrons of those two electron beams. 

\noindent{\bf Supposition 4.} Suppose the light field phase, 
transferred by the captured photon, does not change on the way 
from the dielectric film to the target. 

Then we have for the target point A ($x=0,z$) at a moment $t$ 
(Fig. 1).
\begin{eqnarray}
\label{10}
& & \varphi_0=\varphi_0(0,z,t)=-\omega (t-t_0), \nonumber \\
& & \varphi_1=\varphi_1(0,z,t)=k_xx_1-\omega (t-t_1). 
\end{eqnarray}
Here $t_0$ and $t_1 $ are the corresponding times of flight and 
$(x_1,0)$ is a point where the electrons with energy $E_0+\hbar 
\omega$ started. The light field at the target can be written now 
as 
\begin{eqnarray}
\label{11}
\psi = a e^{i\varphi_0} + b e^{i\varphi_1},
\end{eqnarray}
where the positive constants $a$ and $b$ are determined by the 
currents of the corresponding electron beams. The radiation 
intensity at the target is determined as for the usual light 
interference  
\begin{eqnarray}
\label{12}
I = a^2+b^2+2ab\cos{(\varphi_0-\varphi_1)}. 
\end{eqnarray}
Using the smallness of $\hbar \omega/E_0$, we obtain in the case 
of the collimated incident electron beam
\begin{eqnarray}
\label{13}
\Delta \varphi = \varphi_0-\varphi_1 = \omega (t_0-t_1)-k_xx_1 = 
\nonumber \\ 
= \frac{4\pi z}{\lambda_{b0}}\left [1 - \frac 
{v_0^2}{c^2}(1-n^2\cos ^2\alpha )\right ]. 
\end{eqnarray}
We may consider as in Sect.~\ref{sect:beating2} the more general 
situation when the incident electron beam is focused at the 
distance $r$ before the dielectric film. In this case our model 
gives the following generalization of Eq. (\ref{13})
\begin{eqnarray}
\label{14}
 & & \Delta \varphi =  \frac{4\pi z}{\lambda_{b0}}
\left \{1 - \left (\frac{v_0}{c}\right )^2\left [1-(n \cos{\alpha })^2
\frac{r}{z+r}\right ]\right \}. 
\end{eqnarray} 

This expression differs only by factor 2 from the quantum 
expression (\ref{8}) for the spatial beating phase $\chi $. Both 
the expressions lead to the same spatial period ($\lambda_b/2$) 
for the radiation intensity. We consider this coincidence as a 
serious argument for the acceptance of {\bf Supposition 4} in our 
phenomenological scheme. The supposition looks natural enough 
because a photon in a plane light wave in vacuum also carries a 
constant phase. 

However, the essential difference between our model and the 
quantum treatment is in the initial phase of the spatial 
oscillations of the radiation intensity. The expression (\ref{12}) 
gives the maximum of the Schwarz-Hora radiation intensity at the 
film surface $z$ = 0 unlike the results of the previous quantum 
considerations. Thus our model allows to resolve the problem, 
discussed in the end of Sect.~\ref{sect:beating2}, on the 
constancy of $\lambda_b = 1.70 \AA$. 

\section{CONCLUSIONS}

\label{sect:concl} 

\hspace {3.5mm}
There are several essential contradictions unresolved up to now 
between the theory and the Schwarz experiments. In this work we 
consider in detail one of these contradictions - the problem 
connected with the interpretation of the long-wavelength spatial 
modulation of the Schwarz-Hora radiation. It is shown that the 
experimental values for the long beating wavelength $\lambda_b$ 
can be explained only if the influence of the electron beam 
divergence is taken into account. The model is considered in 
which the incident electron beam is focused at the distance $r$ 
before the dielectric film. The problem of the wavelength 
$\lambda _b$ can be formally resolved if we fix the ratio 
$r/(z+r)$ in the expression (\ref{8}). The additional information 
on the Schwarz experiments or new experiments on the Schwarz-Hora 
effect may clear up the problem of the spatial beating wavelength 
dependence on the film-target distance $z$. 

The physical model, proposed in this work, allows to give the 
more coherent picture of the Schwarz-Hora effect than the 
existing quantum theories: 

1. The model allows to explain the relatively high intensity 
of the Schwarz-Hora radiation. 

2. The model gives the linear dependence of the Schwarz-Hora 
radiation on the electron beam current in agreement with the 
experiment. 

3. The reported by Schwarz strong dependence of the radiation 
intensity on the laser light polarization is a manifestation of 
some spin properties of the electron metastable state. However, 
we do not consider those properties at a present stage of the 
model development. 

4. The model gives the maximum of the radiation intensity at the 
dielectric film surface in agreement with the experiment. 

5. The model allows to reproduce the magnitude and the constancy 
of the spatial beating period reported by Schwarz in 
Ref.\cite{3} if two suppositions about an experimental setup 
are made: (i) the incident electron beam was not strictly 
collimated and was focused at the distance $r$ before the 
dielectric film; (ii) the ratio $r/z$ was kept up fixed in the 
experiments reported in Ref.\cite{3}.

Thus, the Schwarz-Hora effect presents, in our view, the unique 
possibility to observe the metastable states of electrons with 
captured photons and the more detailed experimental study of this 
effect is necessary.

\end{document}